\documentclass[aps,amsmath,amssymb,twocolumn,longbibliography,10pt,pra,superscriptaddress]{revtex4-1}
\usepackage[utf8]{inputenc}
\usepackage[T1]{fontenc}
\usepackage{graphicx}
\usepackage[colorlinks=true,linkcolor=blue,citecolor=red,urlcolor=magenta]{hyperref}
\usepackage{color}
\usepackage{multirow}
\usepackage{tabularx}
\newcolumntype{Y}{>{\centering\arraybackslash}X}

\AtBeginDocument{%
    \newwrite\bibnotes
    \def\bibnotesext{Notes.bib}
    \immediate\openout\bibnotes=\jobname\bibnotesext
    \immediate\write\bibnotes{@CONTROL{REVTEX41Control}}
    \immediate\write\bibnotes{@CONTROL{%
    apsrev41Control,author="08",editor="1",pages="1",title="0",year="1"}}
     \if@filesw
     \immediate\write\@auxout{\string\citation{apsrev41Control}}%
    \fi
}%

\begin{document}

\title{Unconventional pairing in few-fermion systems tuned by external confinement}
\author{Jacek Dobrzyniecki}
\email{Jacek.Dobrzyniecki@fuw.edu.pl}
\affiliation{\mbox{Institute of Physics, Polish Academy of Sciences, Aleja Lotnikow 32/46, PL-02668 Warsaw, Poland}}
\affiliation{ Faculty of Physics, University of Warsaw, ul. Pasteura 5, PL-02093 Warsaw, Poland }
\author{Giuliano Orso}
\affiliation{Universit\'e de Paris, Laboratoire Mat\'eriaux et Ph\'enom\`enes Quantiques, CNRS, F-75013, Paris, France}
\author{Tomasz Sowi\'nski}
\affiliation{\mbox{Institute of Physics, Polish Academy of Sciences, Aleja Lotnikow 32/46, PL-02668 Warsaw, Poland}}

\date{\today}

\begin{abstract}
We study the ground-state properties of a two-component one-dimensional system of a few ultra-cold fermions with attractive interactions. We show that, by ramping up an external potential barrier felt by one of the components, it is possible to induce regions of exotic superfluid phases, characterized by a tunable finite net momentum of the Cooper pair, without changing the overall spin populations. We show that these phases, which are the few-body analogs of the celebrated Fulde-Ferrell-Larkin-Ovchinnikov state, can be distinguished by analyzing a specific two-particle correlation encoded in the noise correlation function. Our theoretical results can be addressed in current experiments with cold atoms confined in spin-selective optical traps.
\end{abstract}

\maketitle

\section{Introduction}

Ultra-cold atoms provide an exceptional experimental platform to simulate condensed matter systems in a controlled way~\cite{stringari,bloch2008}. One of the most spectacular collective phenomena in solids is superconductivity, where electrons  with opposite spin and momenta bind into Cooper pairs, due to an effective attractive interaction mediated by the crystal vibrations.
In the presence of a mismatch  between the two spin populations, generated for instance by an applied Zeeman field, the conventional Bardeen--Cooper--Schrieffer (BCS) pairing mechanism becomes unstable, as some electrons will inevitably end up without partners. The spin-imbalanced system can nevertheless remain superconducting, at the price of adopting a new pairing mechanism harnessing  the excess fermions. A well known example of superconductivity coexisting with a partial spin polarization is  the Fulde-Ferrell-Larkin-Ovchinnikov (FFLO) state~\cite{fulde64,larkin64}, where Cooper pairs condense at a finite momentum.
This implies that the associated order parameter becomes spatially modulated,  with the excess  fermions sitting at the nodes
of the wave, where they are less detrimental to superconductivity.

The modulated phase is currently investigated in different physical systems, including one- and two-dimensional organic superconductors~\cite{KoutroulakisPRL2016},
hybrid structures~\cite{buzdinRMP2005} and quark-gluon plasma~\cite{nardulli04}. Over the last decade atomic Fermi gases have also emerged as a valid alternative to
study this exotic state of matter (for a recent review see~\cite{chevy_2010,feiguin2012,torma1}) for a number of reasons: i) the two spin states correspond to two hyperfine levels, whose populations
are fixed at the beginning of the experiment via a radio-frequency field, without generating vortices; ii) there is no disorder; iii) fermions can be confined in low-dimensional geometries, where the FFLO state is more robust. In particular,  the ground state of one-dimensional (1D) systems with attractive contact interactions is known both theoretically~\cite{yang1,chaohong,Cheng:PRB2018} and numerically~\cite{meisner,batrouni,ueda,mueller1} to be of the FFLO type, for any finite value of the spin imbalance.
The phase diagram of a spin-imbalanced attractive 1D Fermi gas in a harmonic trap has been predicted in ~\cite{orso1,drummond} and experimentally verified in Ref.~\cite{2010-Liao-Nature}. In particular the density profiles of the two spin components develop a two-shell structure, with the central part being a FFLO phase, while the wings are either fully paired or fully polarized, depending on the overall spin polarization. These predictions were also confirmed by DMRG studies of the Fermi Hubbard model~\cite{meisner2010}.
To date, obtaining direct experimental evidence of the modulated phase with cold atoms is still challenging.
Several detection schemes  have been discussed, based on the analysis of collective oscillations
~\cite{2009-Edge-PRL}, the sudden expansion of the gas~\cite{KajalaPRA2011,LuPRL2012,BolechPRL2012}, interaction quenches~\cite{orso2}, noise correlations~\cite{2008-Luscher-PRA,2020-Pecak-PRR}, spectroscopy measurements~\cite{torma3,torma4,roscilde,LutchynPRA2011} and the coupling to a Bose gas~\cite{Singh2020}.

An interesting problem is to investigate the FFLO pairing in 1D Fermi gases in the presence of spin-dependent external potentials, so that the effective Zeeman field,
corresponding to the semi-difference between the local chemical potentials of the two spin components, is no longer uniform throughout the atomic cloud. A natural question then arises: can one tune the external confinement  to induce different superconducting phases in the Fermi gas,  \emph{without} changing the overall spin populations?
To answer this  intriguing question, in this paper we investigate theoretically a 1D spin-1/2 system of a few attractively interacting fermions, confined in a box trap with an additional spin-dependent potential barrier at the trap center. Our main object of interest are the pairing correlations present in this system, which can be analyzed by means of the noise correlation distribution. Depending on the \emph{local} population imbalance, the system behaves either as a BCS superconductor or a FFLO state,  distinguishable by a nonzero center-of-mass-momentum of Cooper pairs. We show that by appropriately tuning the height and width of the potential barrier, it is possible to switch the system between different pairing types, even though the atom numbers of both components remain unchanged.

Although superconducting pairing mechanisms are typically studied for bulk many-body systems, here we explicitly focus on few-body systems far from the thermodynamic limit. In this way our work connects to ongoing experiments on few-body samples \cite{2021-Sowinski-EPL}, such as those being currently undertaken in Jochim's group \cite{2011-Serwane-Science,2012-Zurn-PRL,2013ZurnPRL,2015-Murmann-PRL,2020-Bayha-Nature}.

This work is organized as follows. In Sec. \ref{sec:model}, we describe the model system under study. In Sec. \ref{sec:pair-correlations} we examine the pair correlations that arise in the box trap, without the potential barrier, and show how they can be analyzed through noise correlation distributions. In Sec. \ref{sec:conversions}, we describe the effect of the potential barrier, showing how the dominant net pair momentum changes as the barrier parameters (height and width) are modified. In Sec.\ref{sec:oddnumber}, we describe the particular case where the component split by the potential barrier has an odd number of fermions. Sec. \ref{sec:conclusion} contains the conclusions.

\section{The model}
\label{sec:model}

In this work we consider a one-dimensional system of a few fermions of mass $m$ in two different internal states $\sigma\in\{A,B\}$,  playing the role of effective spins.  Motivated by state-of-the-art experiments with ultra-cold atoms in two hyperfine levels, we assume that the particle numbers $N_A$ and $N_B$ of the two spin components are fixed integers.
We assume that atoms are strongly confined along two orthogonal directions, using for instance a tight two-dimensional optical lattices, so that their motion along these directions reduces to zero-point oscillations. Under this assumption the system behaves kinematically as 1D and is described by a second-quantized Hamiltonian of the form
\begin{align}
\label{eq:ManyBodyHamiltonian}
 \hat{{\cal H}} = &\sum_{\sigma} \int\!\mathrm{d}x \hat{\Psi}_\sigma^\dagger(x) \left(-\frac{\hbar^2}{2m}{{\mathrm{d}^2}\over{\mathrm{d}x^2}}+{\cal V}_\sigma(x) \right) \hat{\Psi}_\sigma(x) \nonumber \\
 &+ g \int\!\mathrm{d}x\,\hat{n}_A(x)\hat{n}_B(x),
\end{align}
where the fermionic field operator $\hat{\Psi}_\sigma(x)$ annihilates a $\sigma$-fermion at position $x$ and obeys the conventional fermionic anticommutation relations, $\{\hat{\Psi}_\sigma(x),\hat{\Psi}_{\sigma '}(x')\} = 0$ and $\{\hat{\Psi}_\sigma(x),\hat{\Psi}^\dagger_{\sigma '}(x')\} = \delta_{\sigma\sigma'}\delta(x-x')$. For convenience, we introduced the single-particle density operators $\hat{n}_\sigma(x)=\hat{\Psi}_\sigma^\dagger(x)\hat{\Psi}_\sigma(x)$. In the following we take the external potential ${\cal V}_\sigma(x)$ as
\begin{equation}
{\cal V}_\sigma(x) = \left\{
\begin{array}{ll}
V_\sigma, & |x| \le D, \\
0,	& D<|x| \le L, \\
\infty, & L < |x|,
\end{array} \right.
\end{equation}
with $V_A=V$ and $V_B$=0, respectively. It means that the particles are confined in an infinite square well of length $2L$, and the component $A$ additionally feels a potential barrier of width $2D$ and height $V$ in the center of the box. From an experimental point of view, a spin-dependent external potential can be achieved {\it e.g.} by using a focused laser beam or magnetic field gradient to induce a spatially localized spin-selective energy shift \cite{2004-Schrader-PRL,2006-Zhang-PRA,2007-Lee-PRL,2011-Weitenberg-Nature}.
The inter-particle interactions are modelled as contact interactions between fermions of opposite species with strength $g$. The interaction strength $g$ is related to the three-dimensional $s$-wave scattering length \cite{1998-Olshanii-PRL,2009-Haller-Science} and can be tuned by magnetic Feshbach resonances \cite{2008-Pethick-Book,2010-Chin-RevModPhys} or by adjusting the confinement in the transverse directions \cite{1998-Olshanii-PRL}.

For convenience, throughout the rest of this paper we employ dimensionless units, {\it i.e.}, we express all energies, lengths, and momenta in units of $\hbar^2/mL^2$, $L$, and $\hbar/L$, respectively. In these units the interaction strength is expressed in units of $\hbar^2/mL$. Without losing the generality of the final conclusions, throughout this paper we set the strength of attractive interactions to $g = -5$. Importantly, we consider 1D systems of few (up to 12) fermions  and assume that  the external potential has fixed spatial size $L$. For this reason, our results cannot be straightforwardly extrapolated to the thermodynamic limit, where the size of the system has to be changed together with the number of particles to keep the average density constant.

To numerically obtain the many-body ground state for the given number of particles and external potential configuration, we first solve the corresponding single-particle eigenproblems for each component separately. Then, we use the lowest-energy eigenorbitals to construct the non-interacting many-body Fock basis $\{ |F_i \rangle \}$. Specifically, each Fock state $|F_i \rangle$ is a product of two Slater determinants of $N_A$ and $N_B$ orbitals, describing the many-body state of $A$ and $B$ component, respectively. The resultant many-body basis, in general, includes all the possible combinations of single-particle orbitals of the $N_A + N_B$ fermions. Since the many-body basis grows exponentially with the number of particles, we limit the basis to Fock states which have a non-interacting energy below a properly chosen value $E_{\mathrm{max}}$, according to the recipe given in \cite{2019-Chrostowski-Acta}. This procedure is based upon the assumption that very high-energy Fock states will be only negligibly represented in the ground-state wave function of the system. Then the many-body Hamiltonian \eqref{eq:ManyBodyHamiltonian} is expressed as a matrix in the basis $\{ |F_i \rangle \}$ and diagonalized using the implicitly restarted Arnoldi method \cite{1998-Lehoucq-Guide}. In this way, the ground state $|G\rangle$ is found as its decomposition in the basis $\{ |F_i \rangle \}$ and used for further calculations. In the end, we confirm that the obtained results do not change quantitatively upon increase of the cutoff energy $E_{\mathrm{max}}$. Thus, the method gives practically exact results, in the sense that the ground-state many-body wave function is known almost exactly, {\it i.e.}, further increase of the Fock basis does not change the results significantly.

\section{Pairing in the trapped system}
\label{sec:pair-correlations}

\begin{figure}
\includegraphics[width=\linewidth]{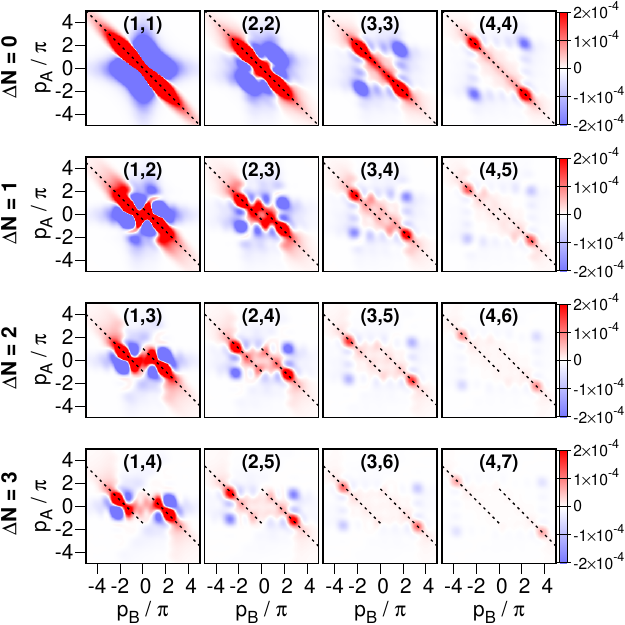}
\caption{The momentum noise correlation $\mathcal{G}(p_A,p_B)$ in the ground state, calculated for systems with $g = -5$, potential barrier $V = 0$, and different particle numbers $(N_A,N_B)$. Each row includes systems with different imbalance $\Delta N = N_B - N_A$. For balanced systems ($\Delta N = 0$), one sees an enhancement of correlations along the line $p_A = -p_B$ (indicated by the dashed line), signalling BCS-like pairing of fermions with opposite momenta. For imbalanced systems, the enhanced correlations instead form along the lines $p_A = -p_B \pm q_0$ (indicated by dashed lines), signalling the creation of FFLO-like pairs with net momentum $\pm q_0$. Momentum is given in units of $\hbar/L$, noise correlation in units of $L^2/\hbar^2$. }
\label{fig:NoiseCorrelationTable}
\end{figure}

For the attractive inter-component interactions ($g < 0$), fermions of opposite species form strongly correlated pairs. It has been shown that these pairs may display many different features of the Cooper pairs known from the theory of superconductivity \cite{2013ZurnPRL,2015-Damico-PRA,2015-Sowinski-EPL,2016HofmannPRA,2020-Lydzba-PRA}. Identifying the type of pairing that arises (\emph{i.e.} pairing with zero or nonzero net pair momentum) requires a detailed knowledge of the superconducting correlation function. It has been previously shown \cite{2008-Luscher-PRA,2020-Pecak-PRR,2020-Rammelmuller-SciPost} that such information can also be obtained from the two-point noise correlation $\mathcal{G}$ between the two components, which is directly experimentally accessible from two-body density measurements \cite{2004-Altman-PRA,2008-Mathey-PRL,2009-Mathey-PRA}. In the momentum domain, the distribution of the noise correlation is given by
\begin{equation}
\mathcal{G}(p_A,p_B) = \langle \hat\pi_A(p_A) \hat\pi_B(p_B) \rangle - \langle \hat\pi_A(p_A) \rangle \langle \hat\pi_B(p_B) \rangle,
\end{equation}
where the momentum density operators $\hat{\pi}_\sigma(p)=\hat{\Psi}^\dagger_\sigma(p) \hat{\Psi}_\sigma(p)$ are expressed straightforwardly in terms of Fourier-transformed field operators $\hat{\Psi}_\sigma(p)=\int \mathrm{d}x \hat{\Psi}_\sigma(x) \exp(-ipx/\hbar)$. The noise correlation is the difference between the two-particle density distribution, and the product of individual single-particle densities. For a non-interacting system, $\mathcal{G}(p_A,p_B)$ is zero everywhere. It means that $\mathcal{G}(p_A,p_B)$ expresses the distribution of correlations forced by inter-particle interactions that cannot be captured by a single-particle description, excluding the spurious correlations that arise from single-particle densities.

To demonstrate that different types of correlated pairs are well-captured by the noise correlations, in Fig.~\ref{fig:NoiseCorrelationTable} we show the distribution $\mathcal{G}(p_A,p_B)$ in the absence of the potential barrier ($V=0$) and different imbalances $\Delta N = N_B - N_A$. For all balanced systems ($\Delta N = 0$), an enhancement of inter-component correlations ($\mathcal{G} > 0$) is visible along the antidiagonal $p_A = -p_B$. It means that the probability of finding a pair of fermions with exactly opposite momenta is enhanced, which is a footprint of the standard Cooper-like pairing mechanism with zero pair momentum. For imbalanced systems ($\Delta N > 0$), the situation is different. The region of enhanced correlations is split into two ridges, located along the two dashed lines, corresponding to net momenta $p_A + p_B = \pm q_0$, with $q_0$ having a nonzero value, a hallmark of FFLO pairing.

The net FFLO pair momentum $q_0$ in the box trap is expected to be equal to the difference $\Delta p_F = p_{FB} - p_{FA}$ between the Fermi momenta $p_{F\sigma} = N_\sigma \pi / 2$ of the two spin components~\cite{noteBox}
\begin{equation}
\label{eq:q0_in_box}
  q_0 = \Delta p_F = \Delta N \pi/2.
\end{equation}
Note that $q_0$ depends only on the population imbalance $\Delta N$ and not on the values $N_A$ and $N_B$ separately, unlike in non-uniform (e.g. harmonically trapped) systems \cite{2020-Pecak-PRR}. The relation \eqref{eq:q0_in_box} is confirmed by Fig.~\ref{fig:NoiseCorrelationTable}: for larger particle numbers, the noise correlation enhancement is concentrated into two clear, narrow maxima  located at $p_A \approx \pm N_A \pi /2, p_B \approx \mp N_B \pi /2$ (red spots); in contrast,
the probability of Cooper pairing between fermions with momenta pointing along the same direction is strongly suppressed (blue spots).
Separately, it is worth noting from Fig.~\ref{fig:NoiseCorrelationTable} that the intensity of the noise correlations diminishes for higher particle numbers, because in 1D systems interactions effects are reduced as the particle density increases \cite{orso1}.  In particular the relevant dimensionless parameter is $\gamma=g m/(\hbar^2 n)$, where $n=(N_A+N_B)/(2L)$ is the total particle density (notice that we have reintroduced all physical units for better clarity).  The parameter $\gamma$ can range from the weakly interacting mean-field regime ($\gamma \ll 1$) to the strongly correlated regime ($\gamma \gg 1$).
Within our units convention the interaction parameter then reduces to $\gamma=2 g/(N_A+N_B)$. For instance for $N_A=N_B=4$  we find $\gamma=1.25$ for $g=-5$, implying that the system is in between the weakly and the strongly interacting regimes.

\begin{figure}
\includegraphics[width=\linewidth]{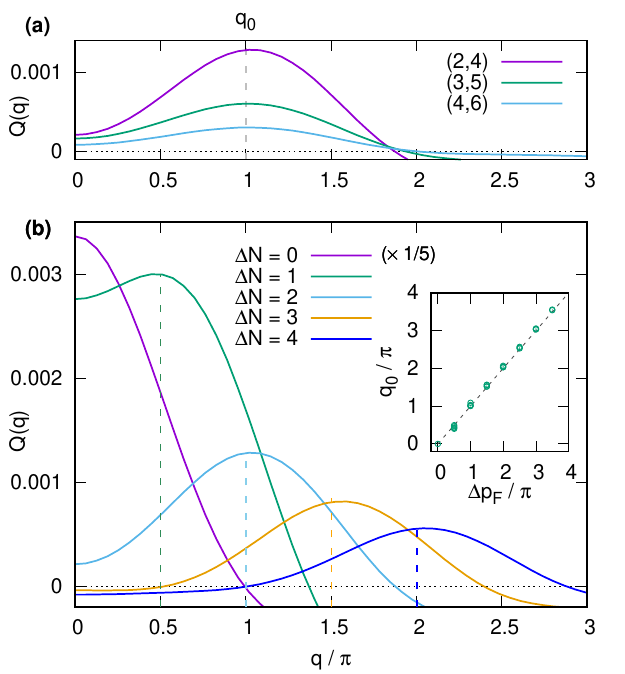}
\caption{(a) The function $\mathcal{Q}(q)$, whose maxima indicate the most probable values of correlated pair momentum $q_0$, for few-body systems with fixed particle imbalance $\Delta N = 2$ and different particle numbers $(N_A,N_B)$. The maximum in each case is present approximately at the value $q = \pi$ (dashed line), equal to the Fermi momentum mismatch $\Delta p_F$. (b) The function $\mathcal{Q}(q)$ in systems with $N_A = 2$ and different values of $N_B = N_A + \Delta N$. Note that the $\Delta N = 0$ curve is scaled by $1/5$. Dashed lines indicate the theoretically predicted momenta $\Delta N \pi / 2$. (Inset) The location of the maximum $q_0$ in $\mathcal{Q}(q)$ (circles) as a function of the theoretically predicted Fermi momenta mismatch $\Delta p_F$. Different points correspond to different particle numbers and imbalances (see the numerical data and explanation in Appendix~\ref{sec:appendix-q0table} for details). The dashed line corresponds to a theoretically predicted exact agreement $q_0 = \Delta p_F$. Momentum is given in units of $\hbar/L$, $\mathcal{Q}(q)$ in units of $L/\hbar$.}
\label{fig:Q_for_imbalances}
\end{figure}

To identify more clearly the most probable net momentum of the pair, $q_0$, we use a method previously proposed in \cite{2020-Pecak-PRR}. It involves integrating the noise correlation with an appropriate filtering function $\mathcal{F}(k)$:
\begin{equation}
\label{eq:Q_definition}
 \mathcal{Q}(q) = \int \mathrm{d}p_A \mathrm{d}p_B\, \mathcal{F}(p_A + p_B - q) \mathcal{G}(p_A,p_B).
\end{equation}
For the filtering function, we choose a simple Gaussian function $\mathcal{F}(k) = (\pi w)^{-1/2} \exp(-k^2/2w^2)$.
The width parameter $w = 0.5$ is of the order of the perpendicular width of the enhanced correlation area. We have checked that the form of $\mathcal{Q}(q)$ is not significantly affected by small adjustments of $w$. Note that $\mathcal{Q}(q) = \mathcal{Q}(-q)$, due to the symmetry of $\mathcal{G}$. Therefore, for simplification, throughout the rest of this paper, we show its values only for positive $q$.
In Fig.~\ref{fig:Q_for_imbalances} we plot the function $\mathcal{Q}(q)$ for systems with different particle numbers.
The momentum $q$ at which the function $\mathcal{Q}$ takes its maximum value can be identified with the most likely net momentum $q_0$ of the Cooper pairs in the system.
Fig.~\ref{fig:Q_for_imbalances}a refers to systems with identical population imbalance $\Delta N = 2$, but varying particle numbers $N_A$ and $N_B$ (they correspond to the third row in Fig.~\ref{fig:NoiseCorrelationTable}). In all these cases, the maximum of $\mathcal{Q}(q)$ falls at the same position $q \simeq \pi$, in agreement with the prediction \eqref{eq:q0_in_box}. Conversely, in Fig.~\ref{fig:Q_for_imbalances}b we show $\mathcal{Q}(q)$ for $N_A = 2$ and different particle imbalances (corresponding to the second column in Fig.~\ref{fig:NoiseCorrelationTable}). The balanced system ($\Delta N = 0$) with BCS pairing is characterized by a clear maximum  at $q \simeq 0$. As the particle imbalance increases, the maximum occurs at increasingly higher momenta, in each case very close to the predicted value \eqref{eq:q0_in_box}, indicated by the vertical dashed lines.
This point is further illustrated in the inset in Fig.~\ref{fig:Q_for_imbalances}, showing the most likely pair momentum $q_0$, defined through the function  $\mathcal{Q}(q)$, versus the Fermi momentum mismatch $\Delta p_F$,
for various particle numbers and imbalances (see the numerical data and explanation in Appendix~\ref{sec:appendix-q0table} for details). All data points fall very close to the dashed straight line, corresponding to $q_0 = \Delta p_F$ as predicted by \eqref{eq:q0_in_box}, confirming that the the most likely net momentum of the Cooper pairs basically coincides with the mismatch between the Fermi momenta across a wide variety of system sizes. Additionally, we have verified that this result does not change qualitatively as the interaction strength is varied, although the intensity of the noise correlation distribution reduces by approaching the weakly interacting regime.

\section{Role of the internal barrier}
\label{sec:conversions}

So far we have assumed that both spin components feel the same external (flat box) potential. Let us now consider the effects of changing the barrier height $V$ felt solely by the component $A$. As $V$ increases, the $A$-fermions are progressively pushed towards the lateral wings until the central region is completely emptied.
In the high-barrier limit, the $A$-fermions effectively experience a symmetrical double-well potential with negligible tunneling between the two wells. Since each separate well has width $1-D$, we see from
 \eqref{eq:q0_in_box} that
the most likely net momentum $q_0$ of Cooper pairs  is given by
\begin{equation}
\label{eq:q0_in_wells}
 q_0 = \Delta N' \pi / (1-D).
\end{equation}
Here $\Delta N' = N'_{B} - N'_{A}$, where $N'_{\sigma}$ is the number of fermions of component $\sigma$ found within a given well. The expected value of $N'_\sigma$ can be determined by integrating the corresponding density profile $n_\sigma(x)=\langle \hat{n}_\sigma(x)\rangle$ over the the well domain.

\subsection{Barrier with varying height}
\begin{figure}
\includegraphics[width=0.9\linewidth]{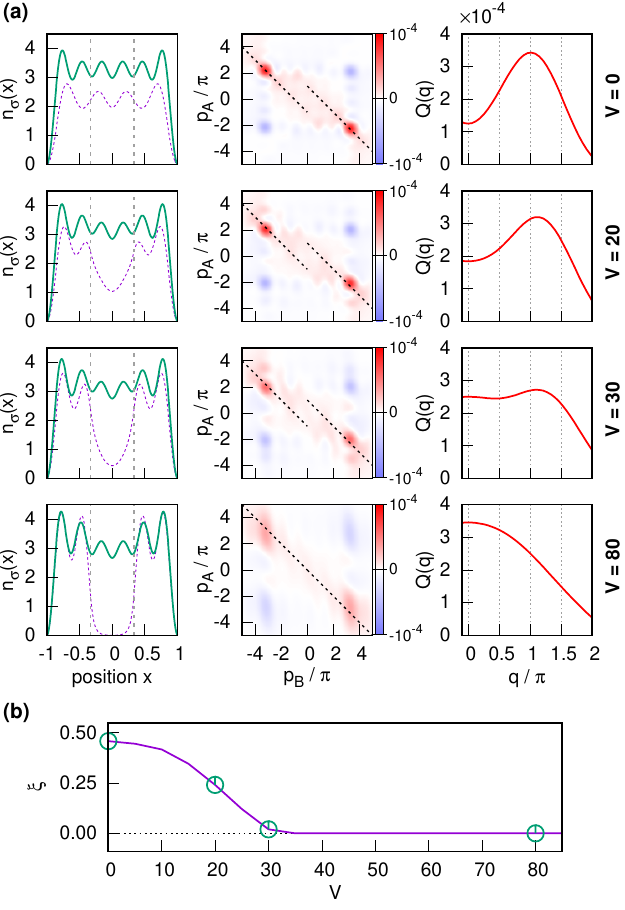}
\caption{The gradual transition of a system with FFLO-like pairing to BCS-like pairing as the potential barrier height $V$ is tuned. The system parameters are $N_A=4,N_B=6,g=-5,D=1/3$. (a) Left column: The one-particle densities $n_\sigma(x)$ in the ground state of the system for increasing $V$, for component $A$ (thin dashed purple lines) and $B$ (thicker solid green lines). Vertical dashed lines indicate the edges of the potential barrier $-D \le x \le D$. As $V$ is increased, the $A$-component fermions are pushed out to the lateral regions, while the distribution of the $B$-fermions remains essentially unchanged. Middle column: Noise correlation distributions $\mathcal{G}(p_A,p_B)$ of the system for increasing $V$. At $V = 0$, clear maxima are visible at $p_A+p_B \approx \pi$. For increasing $V$ these maxima gradually vanish, replaced with more indistinct maxima close to the $p_A = -p_B$ antidiagonal (marked by dashed lines). Right column: The function $\mathcal{Q}(q)$ for increasing $V$. At $V = 0$ a clear maximum is present near $q = \pi$. For increasing $V$, the peak near $q = \pi$ gradually vanishes while a maximum emerges at $q = 0$, indicating the switch from FFLO to BCS pairing. (b) The quantity $\xi$, defined in Eq.~(\ref{eq:chi_value}) and expressing the dominance of non-BCS pairing, as a function of $V$. Here the value of $V$ is smoothly changed over the entire range from $V = 0$ to $V = 85$. Circles indicate the values of $V$ depicted in the above plots. As $V$ is increased and the BCS pairing becomes dominant, $\xi$ decreases to zero. Energy is given in units of $\hbar^2/mL^2$; position in units of $L$; density $n_\sigma$ in units of $1/L$; momenta in units of $\hbar/L$; noise correlation in units of $L^2/\hbar^2$; $\mathcal{Q}(q)$ in units of $L/\hbar$.
}
\label{fig:4+6_case}
\end{figure}

To demonstrate the effect of changing the barrier, let us consider a system with $N_A = 4$ and $N_B = 6$ fermions. In Fig.~\ref{fig:4+6_case}a, we examine its ground state properties for increasing $V$, assuming that the barrier width is fixed to $D = 1/3$. We plot the single-particle densities $n_A(x)$ and $n_B(x)$, along with the corresponding noise correlation distributions $\mathcal{G}(p_A,p_B)$ and the function $\mathcal{Q}(q)$ reflecting the pair momentum distribution. In the homogenous case ($V = 0$, top row in Fig.~\ref{fig:4+6_case}a), both densities (left column) are roughly evenly distributed throughout the box. Due to the population imbalance, the noise correlations in the system support the FFLO-like pairing. The pair momentum calculated from \eqref{eq:q0_in_box} is $q_0= \pi$, as seen from the locations of the positive maxima in $\mathcal{G}$ (middle column).

As $V$ increases, the component $A$ is gradually pushed out of the barrier region. Due to the symmetry of the system, the density $n_A(x)$ is evenly split between the two side regions. Meanwhile, the density $n_B(x)$ is essentially unchanged (except for the slight modifications due to the attractive interaction). The particular choice $D = 1/3$ means that approximately one-third of the $B$ population (two fermions) is located within each lateral region. As a result, for high $V$, the population within the lateral regions becomes balanced. This is additionally supported by the fact that densities $n_A(x)$ and $n_B(x)$ in these regions become almost identical. Thus, in this regime, the pairs created within the lateral regions are standard BCS Cooper-like pairs with zero net momentum. This phenomenological reasoning is supported by the noise correlations -- for the large $V$ case (last row) the maxima are found close to the anti-diagonal $p_A = -p_B$. The maxima become gradually more indistinct and stretched along the $p_A$ direction as $V$ increases, which can be explained by the increasing uncertainty of $p_A$ as the $A$ fermions are squeezed into a smaller space.

The transition between FFLO and BCS pairing can be explained in greater detail by inspection of the function $\mathcal{Q}(q)$ (right column in Fig.~\ref{fig:4+6_case}a). For $V = 0$, $\mathcal{Q}(q)$ displays a maximum at $q_0 = \pi$, exactly as predicted for this imbalanced system from the difference in Fermi momenta. As $V$ increases, this maximum gradually vanishes, while simultaneously another maximum emerges at $q_0 = 0$. In particular, for $V = 30$ one can distinguish two separate maxima at the two locations. This indicates that the change between the FFLO and BCS pairings is not a gradual decrease of $q_0$, but rather a direct switch between two distinct pairing mechanisms. A separate effect is that for small barrier heights $V$, the maximum at $q_0 = \pi$ shifts towards slightly larger momenta, which can be explained by the fact that the momentum of $A$-fermions slightly increases due to the higher external potential energy.

From an experimental point of view, it is useful to define an additional measurable quantity that indicates whether the ground state of the system displays BCS or FFLO pairing. For this purpose, we define the dimensionless quantity
\begin{equation}
\label{eq:chi_value}
 \xi = \frac {\int [\mathcal{Q}(q) - Q(0)] \theta[\mathcal{Q}(q) - Q(0)] \mathrm{d}q} {\int \mathcal{Q}(q) \theta[\mathcal{Q}(q)] \mathrm{d}q},
\end{equation}
where $\theta(z)$ is the Heaviside step function. If the maximum of $\mathcal{Q}(q)$ is located at $q=0$, $\xi$ is exactly zero, while if the maximum falls at any other position then $\xi > 0$. The value of $\xi$ can therefore be interpreted as an indicator for FFLO pairing.
In Fig.~\ref{fig:4+6_case}b we show in more details the dependence of $\xi$ on the barrier height for the $N_A=4,N_B=6$ system. As $V$ increases, $\xi$ gradually diminishes and eventually vanishes around $V\simeq 35$, signaling the transition towards the standard BCS pairing.

\subsection{Tuning the barrier width}

\begin{figure}
\includegraphics[width=0.83\linewidth]{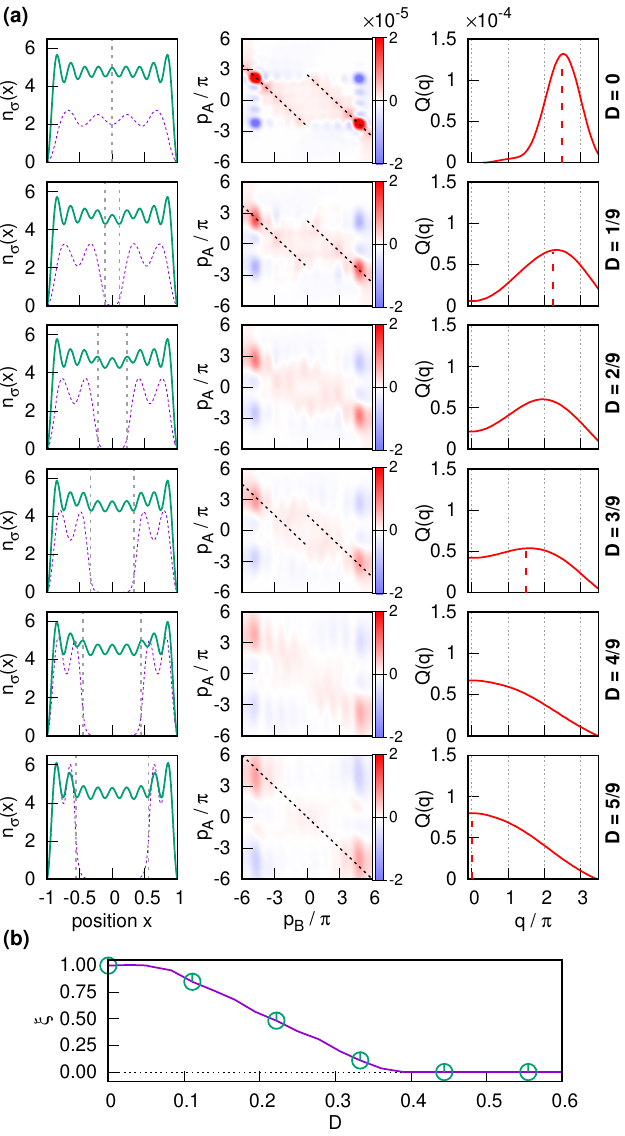}
\caption{The gradual change of the FFLO momentum as the barrier width parameter $D$ is tuned. The system parameters are $N_A=4,N_B=9,g=-5,V=150$. (a) Left column: The one-particle densities $n_\sigma(x)$ in the ground state of the system for increasing $D$. Vertical dashed lines indicate the edges of the potential barrier $-D \le x \le D$. Middle column: Noise correlation distributions $\mathcal{G}(p_A,p_B)$ of the system for increasing $D$. As $D$ increases, the maxima become broader and more indistinct, owing to the increased uncertainty of momentum $p_A$. Dashed lines indicate the predicted locations of maxima $p_A + p_B = q_0$ for given $D$. Right column: The function $\mathcal{Q}(q)$ for increasing $V$. For the values of $D$ where population $N'_{B}$ is well-defined, dashed lines indicate the predicted location of the preferred net momentum $q_0$, calculated from the difference of Fermi momenta. Particularly, at $D = 1/9$ and $D=3/9$, where the population of $A$ and $B$ in the lateral regions is integer but different, the system exhibits a FFLO pairing with a changing net momentum. Meanwhile, for $D = 5/9$, the populations become not only integer but also equal, thus the BCS pairing with $q_0 = 0$ dominates. On the other hand, for widths not supporting integer $N'_B$, particle number in lateral region has significant variance and the pairing correlation visible in the noise cannot be easily associated with a single pairing mechanism (see the main text for details). (b) Value of the quantity $\xi$ as a function of $D$, for $D$ smoothly varied between $D = 0$ and $D = 0.6$. Circles indicate the values of $D$ depicted in the above plots. It is seen that the BCS pairing becomes dominant for $D \gtrsim 0.4$. Position and barrier width is given in units of $L$; density $n_\sigma$ in units of $1/L$; momenta in units of $\hbar/L$; noise correlation in units of $L^2/\hbar^2$; $\mathcal{Q}(q)$ in units of $L/\hbar$.}
\label{fig:4+9_case}
\end{figure}

The FFLO pair momentum $q_0$ can also be controlled by tuning the width of the barrier. To demonstrate this, in Fig.~\ref{fig:4+9_case}a we show results for a system with $N_A = 4$, $N_B = 9$ particles obtained by progressively increasing the barrier width $D$, assuming a fixed barrier height $V = 150$. The latter choice ensures that the middle region is nearly emptied of $A$-fermions. The single-particle densities $n_\sigma(x)$ (left column) give an approximate view of the changing population difference in the lateral regions.

The first row shows the case of a system without a potential barrier. As seen from the noise correlation distribution $\mathcal{G}(p_A,p_B)$ and the function $\mathcal{Q}(q)$ (middle and right column in Fig.~\ref{fig:4+9_case}a, respectively), in this case FFLO-like pairs are formed with a nonzero momentum, $q_0 = 5\pi/2$. The subsequent rows show the effect of changing $D$ to different values. At the values $D = 1/9,3/9,5/9$, the expected value of $N'_{B}$ is a clearly defined integer number (four, three, and two, respectively). Meanwhile, $N'_{A}$ remains close to two in all cases. For these values of $D$, the noise correlation distribution and the function $\mathcal{Q}(q)$ show that the most probable pair momentum $q_0$ changes to the value predicted from \eqref{eq:q0_in_wells} indicated by vertical dashed lines ($q_0 = 9 \pi /4$, $3 \pi / 2$, and $0$, respectively). Additionally, we show the cases of intermediate widths $(D=2/9, D=4/9)$ which lie in between the above values. In such cases the value $N'_{B}$ is non-integer, implying that the ground state is a superposition of different quantum states with different numbers of $B$-particles in the left and right wells. For instance for $D=2/9$ there could be four $B$-fermions in the left well and three $B$-fermions in the right one or the other way round.

These results clearly show that adjusting the barrier widths allows tuning the FFLO momentum $q_0$, as well as switching from FFLO pairing to BCS pairing. This point is further illustrated in Fig.~\ref{fig:4+9_case}b, showing the behavior of $\xi$ as a function of $D$ when $D$ is smoothly varied. In particular $\xi$ decreases monotonically as $D$ increases, and ultimately vanishes around $D\simeq 0.4$, marking the dominance of BCS pairing.

\section{Odd particle number $N_A$}
\label{sec:oddnumber}

\begin{figure}
\includegraphics[width=0.95\linewidth]{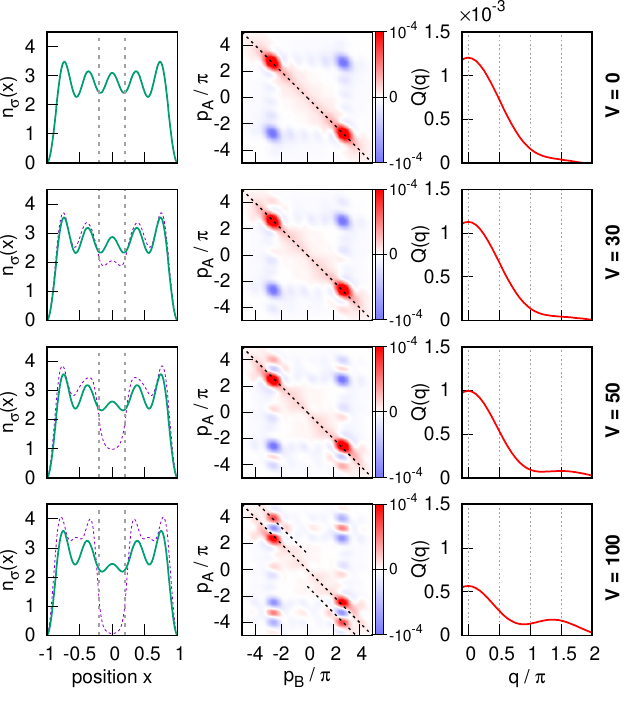}
\caption{The gradual change of a balanced system with odd $N_A$ as the potential barrier height $V$ is tuned. The system parameters are $N_A=5,N_B=5,g=-5,D=1/5$. (a) Left column: The one-particle densities $n_\sigma(x)$ in the ground state of the system for increasing $V$. Vertical dashed lines indicate the edges of the potential barrier $-D \le x \le D$. For $V = 0$ the two densities are exactly identical. For a high potential barrier $V$, the $A$ fermions are pushed out to the lateral regions. As the number of $A$ fermions is odd, the density $n_A(x)$ within each well represents the contributions from density profiles corresponding to two or three fermions. Middle column: Noise correlation distributions $\mathcal{G}(p_A,p_B)$ of the system for increasing $V$. For a high barrier, two distinct maxima can be distinguished at $p_1+p_2 = 0$ and $p_1+p_2 = \pm 5\pi/4$ (marked by dashed lines). Right column: The function $\mathcal{Q}(q)$ for increasing $V$. For a very high barrier, two maxima can be distinguished at $q_0 = 0$ and $q_0 = 5\pi/4$. Energy is given in units of $\hbar^2/mL^2$; position in units of $L$; density $n_\sigma$ in units of $1/L$; momenta in units of $\hbar/L$; noise correlation in units of $L^2/\hbar^2$; $\mathcal{Q}(q)$ in units of $L/\hbar$.}
\label{fig:5+5_case}
\end{figure}

So far we have considered systems with an even particle number $N_A$. In those cases, the introduction of the potential barrier leads (on average) to an equal distribution of $A$-fermions in the two lateral regions. A more complicated situation occurs for systems with odd $N_A$, as the presence of the barrier leads to unequal numbers of $A$-fermions in the two wells.
 As a consequence, the pair momentum $q_0$ can also take distinct values in the two lateral regions.
To demonstrate this, we focus on the balanced case with $N_A = N_B = 5$ particles. The single-particle densities, noise correlation and the function $\mathcal{Q}(q)$ for increasing values of $V$ and fixed $D=1/5$ are shown in Fig.~\ref{fig:5+5_case}. For $V = 0$, the population of both components is exactly balanced and the system is characterized by BCS-like pairing with net pair momentum $q_0 = 0$.
For very high $V$ the most probable distribution is that of two $A$-fermions in one well, and three $A$-fermions in the other. In contrast, the expected number of $B$-fermions in either of the two wells is two, due to the chosen barrier width.
Thus the imbalance $\Delta N'$ is different in both wells, resulting in a dominance of two different values of net pair momentum, $q_0 = 0$ and $q_0 = 5\pi/4$, corresponding to a BCS and a FFLO state, respectively.
This effect is indeed visible in the noise correlation $\mathcal{G}(p_A,p_B)$ and the function $\mathcal{Q}(q)$, where distinct separate maxima are visible at the predicted values of $q$ (last row in Fig.~\ref{fig:5+5_case}).

\begin{figure}
\includegraphics[width=0.95\linewidth]{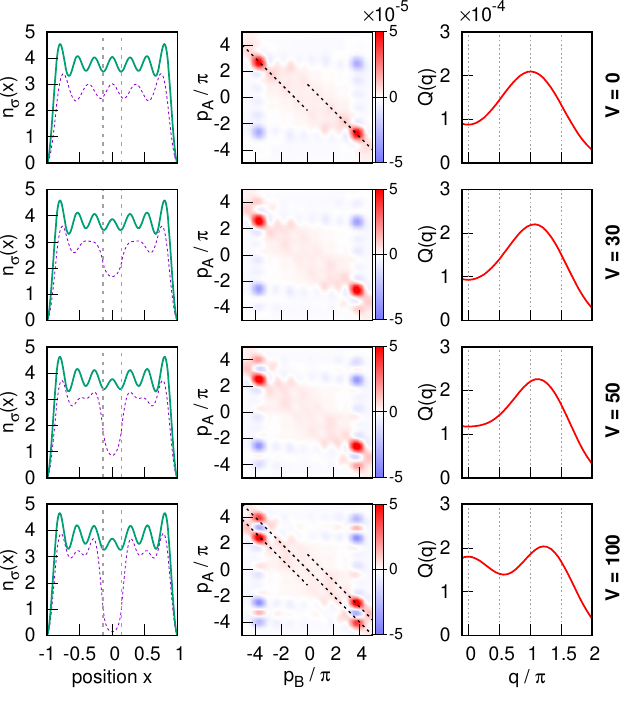}
\caption{The gradual change of an imbalanced system with odd $N_A$ as the potential barrier height $V$ is tuned. The system parameters are $N_A=5,N_B=7,g=-5,D=1/7$. (a) Left column: The one-particle densities $n_\sigma(x)$ in the ground state of the system for increasing $V$. Vertical dashed lines indicate the edges of the potential barrier $-D \le x \le D$. Middle column: Noise correlation distributions $\mathcal{G}(p_A,p_B)$ of the system for increasing $V$. At $V = 0$, clear maxima are visible at $p_1+p_2 = \pi$ (marked by dashed lines). For a high barrier, two distinct maxima can be distinguished at $p_1+p_2 = 0$ and $p_1+p_2 = \pm 7\pi/6$. Right column: The function $\mathcal{Q}(q)$ for increasing $V$. At $V = 0$, a single peak is present at $q_0 = \pi$ while for a very high barrier two peaks can be distinguished at $q_0 = 0$ and $q_0 = 7\pi/6$. Energy is given in units of $\hbar^2/mL^2$; position in units of $L$; density $n_\sigma$ in units of $1/L$; momenta in units of $\hbar/L$; noise correlation in units of $L^2/\hbar^2$; $\mathcal{Q}(q)$ in units of $L/\hbar$.}
\label{fig:5+7_case}
\end{figure}

For completeness, we also investigate the imbalanced case $N_A = 5$, $N_B = 7$, where the system exhibits FFLO pairing already at $V = 0$. In Fig.~\ref{fig:5+7_case} we show the single-particle densities, noise correlations, and the function $\mathcal{Q}(q)$ in this imbalanced system for increasing values of $V$ and a fixed barrier width $D = 1/7$. At $V = 0$, the system exhibits a FFLO pairing with a preferred net momentum $q_0 = \pi$. For very high $V$ the most probable distribution is again that of two $A$-fermions in one well, and three $A$-fermions in the other,  while the expected number of $B$ fermions in each well is three. In this regime $\mathcal{Q}(q)$ exhibits two separate peaks, $q_0 = 0$ and $q_0 = 7\pi/6$. Therefore in the imbalanced odd-$N_A$ case, where the system already exhibits FFLO pairing in the absence of a barrier, tuning the barrier height $V$ from zero leads to two distinct effects: \emph{i)} it can introduce an additional BCS phase in one of the two wells, and \emph{ii)} it can modify the value of the  FFLO momentum $q_0$.

At this point it is valuable to clarify a general structure of the many-body ground state in the odd $N_A$ case. Due to the spatial left-right symmetry, the many-body ground state of the system can be expressed as a general superposition $|G\rangle = (|L\rangle + |R\rangle)/\sqrt{2}$, where $|L\rangle$ and $|R\rangle$ represent many-body states describing configurations with the extra $A$ fermion placed in the left and right well, respectively. This means that it is not possible to distinguish whether it is the left or the right well that contributes to a particular pairing mechanism. In the limit of very large $V$, the ground state becomes nearly degenerate with the first excited many-body state of the system $|G'\rangle = (|L\rangle - |R\rangle)/\sqrt{2}$, having an essentially different single-particle momentum distribution. As a result, in experimental practice, a system prepared in this regime may end up in either one or any superposition of states $|L\rangle, |R\rangle$.

\begin{figure}
\includegraphics[width=0.95\linewidth]{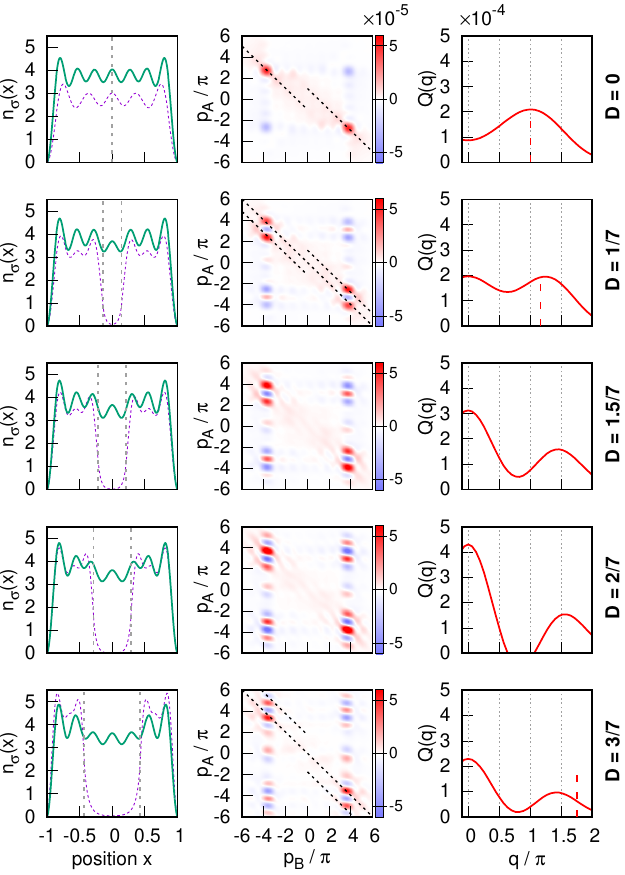}
\caption{The gradual change of a system with odd $N_A$ as the potential barrier width parameter $D$ is tuned. The system parameters are $N_A=5,N_B=7,g=-5,V=120$. (a) Left column: The one-particle densities $n_\sigma(x)$ in the ground state of the system for increasing $V$. Vertical dashed lines indicate the edges of the potential barrier $-D \le x \le D$. Middle column: Noise correlation distributions $\mathcal{G}(p_A,p_B)$ of the system for increasing $V$. For $D=0$, as well as $D=1/7, D=3/7$ for which the population $N'_B$ is well defined, dashed lines indicate the predicted values of FFLO pair momenta $p_A + p_B = q_0$. Right column: The function $\mathcal{Q}(q)$ for increasing $V$. Dashed vertical lines indicate the predicted locations of $q_0$. Energy is given in units of $\hbar^2/mL^2$; position in units of $L$; density $n_\sigma$ in units of $1/L$; momenta in units of $\hbar/L$; noise correlation in units of $L^2/\hbar^2$; $\mathcal{Q}(q)$ in units of $L/\hbar$.}
\label{fig:5+7_case_D}
\end{figure}

Let us finally investigate the behavior of the system with $N_A = 5, N_B = 7$ as the barrier width $D$ is progressively increased from zero, for a fixed barrier height $V = 120$. The obtained results are displayed in  Fig.~\ref{fig:5+7_case_D}. The distributions $\mathcal{Q}(q)$ for no barrier ($D = 0$) and for $D = 1/7$ are as shown previously in Fig.~\ref{fig:5+7_case}, with the preferred FFLO momentum being $q_0 = \pi$ and $q_0 = 7\pi/6$, respectively (as marked by vertical dashed lines in the third column). For $D = 3/7$, the expected value of $N_B'$ is a clearly defined integer number ($N_B' = 2$) and the two momenta are expected to fall at $q_0 = 0$ and $q_0 = 7\pi/4$. However, it can be seen that the position of the maximum in $\mathcal{Q}(q)$ visibly deviates from the predicted $q_0$ in this case, although it is reasonably close to the predicted value. We additionally show the results for two intermediate values of $D$ ($D = 1.5/7, D=2/7$). Since in these cases the expected value $N_B'$ is not a clearly defined integer, the distributions $\mathcal{Q}(q)$ in this case are not straightforward to describe.

\section{Conclusion}
\label{sec:conclusion}

Few-body cold atom systems  represent an intriguing platform to study pairing phenomena.
Here we have investigated a one-dimensional system of few attractively interacting spin-1/2 fermions confined in a flat box trap, through the numerical study of the ground state density profiles and noise correlations. We show that by ramping up a central potential barrier felt by one of the two components, and thus restricting pair formation to regions outside the barrier, the system can undergo different pairing mechanisms without changing the overall spin populations.
Specifically, solely by adjusting the barrier height and width, the particles in the two wells can form either BCS-like pairs with zero center-of-mass momentum, or FFLO-like pairs with a tunable finite momentum.
Moreover, we found that for odd particle numbers both BCS and FFLO type correlations can coexist in different spatial regions of the system, even in the absence of an overall spin imbalance, provided the barrier parameters are tailored appropriately.

Our theoretical results are relevant for current experiments using quasi-1D atomic samples with few particles per tube. In these systems the noise correlations have been
measured with great accuracy, while the spin-dependent external potential can be tailored with state-of-the-art optical techniques. Our work therefore provides a promising route to investigate the FFLO pairing mechanism starting from experiments with small-size cold atoms systems acting as superconducting grains. The present study can also be generalized
to higher dimensions, assuming that the transverse confinement is reduced so that some of the excited single-particle states in the transverse directions become populated. For instance, one can study how the intensity of the noise correlation function depends on the numbers of particles in the system, for a fixed system size; we expect that, contrary to the 1D case discussed here, the intensity increases as the atom density increases, because interaction effects become stronger. It would also be interesting to understand whether the signatures of FFLO pairing in the noise correlation function remain visible in higher dimensions, where the exotic superfluid is known to be less robust.

Another promising direction, which directly connects with $\pi$-phases~\cite{2010-Zapata-PRL} and hybrid Josephson junctions~\cite{buzdinRMP2005},
 is to consider that both left and right side of the barrier represent bulk 1D systems with uniform densities (far from the barrier region).  This corresponds to taking the thermodynamic limit $N_\sigma\rightarrow +\infty$ and $L \rightarrow +\infty$, with $N_\sigma/L$ and the barrier width $D$ being finite. It is worth stressing that such a limit is out of reach for the method used in this work,  because the computational effort required to obtain convergent results grows too fast with system size and particle numbers. Other numerical approaches are more suitable to extract the ground state properties for bulk systems, including Quantum Monte Carlo or the Density Matrix Renormalization Group (DMRG). We leave this program for future studies.

All numerical data presented in this paper is freely available online \cite{zenodo-dataset}.

\section{Acknowledgments}

J.D. and T.S. acknowledge financial support from the (Polish) National Science Center (Grant No. 2016/22/E/ST2/00555). G. O. acknowledges financial support from ANR (Grant
SpiFBox) and from DIM Sirteq (Grant EML 19002465 1DFG).

\appendix

\section{Net pair momenta for different particle imbalances}
\label{sec:appendix-q0table}

In the table \ref{tbl:q0_values} we show the data used in the inset of Fig.~\ref{fig:Q_for_imbalances}.

\renewcommand{\arraystretch}{1.14} 
\begin{table}[h!]
\begin{tabularx}{.35\textwidth}{ |Y Y|Y||Y|Y| }
\hline
$N_A$ & $N_B$ & $\Delta N$ & $\Delta p_F / \pi$ & $q_0 / \pi$ \\
 \hline
 \hline
1 & 2 & \multirow{5}{*}{1} & \multirow{5}{*}{0.5} & 0.50 \\
 \cline{1-2}\cline{5-5}
2 & 3 & & & 0.47 \\
 \cline{1-2}\cline{5-5}
3 & 4 & & & 0.44 \\
 \cline{1-2}\cline{5-5}
4 & 5 & & & 0.42 \\
 \cline{1-2}\cline{5-5}
5 & 6 & & & 0.40 \\
 \hline
 \hline
1 & 3 & \multirow{5}{*}{2} & \multirow{5}{*}{1} & 1.09 \\
 \cline{1-2}\cline{5-5}
2 & 4 & & & 1.03 \\
 \cline{1-2}\cline{5-5}
3 & 5 & & & 1.02 \\
 \cline{1-2}\cline{5-5}
4 & 6 & & & 1.01 \\
 \cline{1-2}\cline{5-5}
5 & 7 & & & 1.01 \\
 \hline
 \hline
1 & 4 & \multirow{4}{*}{3} & \multirow{4}{*}{1.5} & 1.57 \\
 \cline{1-2}\cline{5-5}
2 & 5 & & & 1.56 \\
 \cline{1-2}\cline{5-5}
3 & 6 & & & 1.53 \\
 \cline{1-2}\cline{5-5}
4 & 7 & & & 1.52 \\
 \hline
 \hline
1 & 5 & \multirow{4}{*}{4} & \multirow{4}{*}{2} & 2.07 \\
 \cline{1-2}\cline{5-5}
2 & 6 & & & 2.05 \\
 \cline{1-2}\cline{5-5}
3 & 7 & & & 2.04 \\
 \cline{1-2}\cline{5-5}
4 & 8 & & & 2.03 \\
 \hline
 \hline
1 & 6 & \multirow{3}{*}{5} & \multirow{3}{*}{2.5} & 2.57 \\
 \cline{1-2}\cline{5-5}
2 & 7 & & & 2.55 \\
 \cline{1-2}\cline{5-5}
3 & 8 & & & 2.53 \\
 \hline
 \hline
1 & 7 & \multirow{3}{*}{6} & \multirow{3}{*}{3} & 3.06 \\
 \cline{1-2}\cline{5-5}
2 & 8 & & & 3.04 \\
 \cline{1-2}\cline{5-5}
3 & 9 & & & 3.03 \\
 \hline
 \hline
1 & 8 & \multirow{2}{*}{7} & \multirow{2}{*}{3.5} & 3.56 \\
 \cline{1-2}\cline{5-5}
2 & 9 & & & 3.54 \\
 \hline
 \hline
1 & 9 & \multirow{2}{*}{8} & \multirow{2}{*}{4} & 4.06 \\
 \cline{1-2}\cline{5-5}
2 & 10 & & & 4.04 \\
 \hline
\end{tabularx}
\caption{The most probable net pair momentum $q_0$ for systems with different particle numbers $N_A,N_B$ and imbalances $\Delta N = N_B - N_A$. The system parameters are $g = -5$ and $V = 0$. The pair momentum $q_0$ is found as the location of the maximum of function $\mathcal{Q}(q)$, defined in \eqref{eq:Q_definition}. It is compared to the theoretically predicted Fermi momentum difference $\Delta p_F = \Delta N \pi / 2$. It is seen that $q_0 \approx \Delta p_F$ in all cases, as predicted. Not shown are the entries for $N_A = N_B$, for which in all tested cases $q_0 = \Delta p_F = 0$.}
\label{tbl:q0_values}
\end{table}

\bibliography{_Biblio}

\begin{thebibliography}{60}
\bibitem [{not()}]{noteBox}%
  \BibitemOpen
  \href@noop {} {}\bibinfo {note} {We recall that in our units the system size
  is equal to $2$.}\BibitemShut {Stop}%
\end{thebibliography}

\end{document}